\documentclass{emulateapj}
\usepackage{verbatim,graphicx,natbib}
\shortauthors{Cowan, Holder \& Kaib}
\shorttitle{Searching for Planet Nine in the Millimeter}

\def\aap{A\&A}
\def\apj{ApJ}

\def\mnras{MNRAS}
\def\aj{AJ}
\def\icarus{Icarus}

\def\nat{Nature}

\begin{document}

\title{Cosmologists in Search of Planet Nine: the Case for CMB Experiments}
\author{Nicolas B. Cowan\altaffilmark{1,2,3,4}, Gil Holder\altaffilmark{2,3}, Nathan A. Kaib\altaffilmark{5}}

\altaffiltext{1}{Department of Earth \& Planetary Sciences, McGill University, 3450 rue University, Montr\'eal, QC,  H3A 0E8, Canada\\ email: nicolas.cowan@mcgill.ca} 
\altaffiltext{2}{Department of Physics, McGill University, 3600 rue University, Montr\'eal, QC,  H3A 2T8, Canada}
\altaffiltext{3}{McGill Space Institute}
\altaffiltext{4}{Institut de recherche sur les exoplan\`etes}
\altaffiltext{5}{Homer L.\ Dodge Department of Physics \& Astronomy, University of Oklahoma, 440 W.\ Brooks St., Norman, OK 73019, USA}

\begin{abstract}
Cosmology experiments at mm-wavelengths can detect Planet Nine if it is the size of Neptune, has an effective temperature of 40~K, and is 700~AU from the Sun.  It would appear as a $\sim$30~mJy source at 1~mm with an annual parallax of $\sim$5~arcmin.   The challenge is to distinguish it from the approximately 4000 foreground asteroids brighter than 30~mJy. Fortunately, these asteroids are known to the Minor Planet Center and can be identified because they move across a resolution element in a matter of hours, orders of magnitude faster than Planet Nine.   If Planet Nine is smaller, colder, and/or more distant than expected, then it could be as faint as 1~mJy at 1~mm.  There are roughly $10^6$ asteroids this bright and many are unknown, making current cosmology experiments confusion limited for moving sources.  Nonetheless, it may still be possible to find the proverbial needle in the haystack using a matched filter. This would require mm telescopes with high angular resolution and high sensitivity in order to alleviate confusion and to enable the identification of moving sources with relatively short time baselines.  Regardless of its mm flux density, searching for Planet Nine would require frequent radio measurements for large swaths of the sky, including the ecliptic and Galactic plane.  Even if Planet Nine had already been detected by other means, measuring its mm-flux would constrain its internal energy budget, and therefore help resolve the mystery of Uranus and Neptune, which have vastly different internal heat.  
\end{abstract}

\section{Introduction}
Evidence is mounting for an unseen planet in the Oort cloud \citep{Trujillo_2014, Fuente_Marcos_2014, Fuente_Marcos_2015, Batygin_2016}.  This so-called Planet Nine (a.k.a.\ Planet X, Bowie) betrays its presence by the anisotropic orbits of Sedna-like objects. Astronomers are refining the predictions of Planet Nine's location based on orbital dynamics \citep{Fuente_Marcos_2016}, constraints on gravitational perturbations \citep{Fienga_2016}, and constraints on reflected light \citep{Brown_2016}. 

Any object with internal power exceeding incident starlight emits more energy at thermal wavelengths than it reflects.  This is likely the case for the purported Planet Nine, a $\gtrsim$10$M_\oplus$ planet at $\sim$700~AU \citep[][]{Batygin_2016}.  Even if a planet has negligible internal heat, thermal emission exceeds reflected light if its Bond albedo is less than 50\%, as for all Solar System planets other than Venus.  These simple facts motivate the search for Planet Nine's blackbody emission.  

In this Letter, we propose that cosmology experiments at mm wavelengths are a viable way to search for---and characterize---Planet Nine.  In Section~2 we estimate the temperature and thermal radiation of Planet Nine and in Section~3 we estimate its motion through the sky. In Section~4 we show that the point-source sensitivity and angular resolution of many mm cosmology experiments are sufficient to detect and identify Planet Nine.  We consider false positives in Section~5, and discuss our findings in Section~6.

\section{Thermal Flux from Planet Nine}
\cite{Batygin_2016} suggest that Planet Nine is Neptune-mass and at a distance of approximately 700~AU.  We use the energy budgets of Neptune and Uranus to place constraints on the internal heat of Planet Nine.

The effective temperature of a planet, $T_{\rm eff}$, is the sum of internal power and absorbed solar power: $T_{\rm eff}^4 = T_{\rm iso}^4 + T_{\rm rad}^4$, where $T_{\rm iso}$ is the temperature the planet would have in isolation, and $T_{\rm rad}= (1-A)^{\frac{1}{4}} T_{\rm equ}$ is the temperature of the planet due solely to stellar radiation ($A$ is Bond albedo and $T_{\rm equ}$ is equilibrium temperature).

We may therefore solve for the temperature a planet would have in the absence of an external energy source:  
\begin{equation}
T_{\rm iso} = \left(\frac{\zeta-1}{\zeta} \right)^{\frac{1}{4}} T_{\rm eff},
\end{equation}
where $\zeta$ is the ratio of total energy output to stellar energy input: $\zeta \equiv (T_{\rm eff}/T_{\rm rad})^4$.  

Table 1.3 of \cite{Taylor_2010} lists the values of $\zeta$ for Uranus and Neptune: 1.06 and 2.52, respectively.  In other words, Neptune has greater internal heat than Uranus.  \cite{Hildebrand_1985} measured the effective temperatures of Uranus and Neptune: $T_{\rm eff}= 58$~K and 60~K, respectively. 
This allows us to estimate the isolation temperature of Uranus and Neptune to be $T_{\rm iso} = 28$~K and 53~K, respectively.   We adopt this as a plausible range for Planet Nine.

We next estimate the solar heating at 700~AU.  We find that Planet Nine would have a temperature of $\sim$10~K, in the absence of internal heat and assuming that Planet Nine has the same Bond albedo as Uranus and Neptune (0.30 and 0.29, respectively).  In any case, even in the zero-albedo limit, the energy budget of Planet Nine is dominated by internal heat.  The exact albedo is therefore irrelevant.

Since $T_{\rm iso}>T_{\rm rad}$ for Planet Nine (and therefore $T_{\rm iso}^4\gg T_{\rm rad}^4$), we expect that its effective temperature will be approximately given by its internal temperature: $T_{\rm eff} \gtrsim T_{\rm iso}$.  We therefore expect Planet Nine to have an effective temperature of $T_{\rm eff} = 28$--53~K, corresponding to a blackbody peak of 50--100~micron. Anyone wishing to detect the thermal emission of this planet had better look in the far-IR and longward.

\cite{Hildebrand_1985} presented far-infrared fluxes of the Solar System giant planets. Their Figure~8 shows the flux of Uranus and Neptune from 30 to 1000~micron.  They found that Uranus had a flux density of $F_\nu= \{200, 70\}$~Jy and Neptune has a flux density of $F_\nu= \{60, 25\}$~Jy at $\lambda= \{500, 1000\}$~micron.

Scaling from Neptune, we obtain a 1~mm flux density of:
\begin{equation}
F_{\rm 300GHz} = 30~{\rm mJy} \left(\frac{T_{\rm eff}}{40~{\rm K}}\right) \left(\frac{R_p}{R_{\rm N}}\right)^2 \left(\frac{d}{700~{\rm AU}}\right) ^{-2},
\end{equation}
where $R_p$ is the radius of Planet Nine, $R_{\rm N}$ is the radius of Neptune, and $d$ is Planet Nine's current heliocentric ($\sim$geocentric) distance. This flux is consistent with the planetary evolution models of \cite{Linder_2016} and \cite{Ginzburg_2016}.

Most cosmology experiments designed to map the cosmic microwave background (CMB) have channels at/near 1~mm (300~GHz), but they are typically optimized for 2~mm radiation (150~GHz).  Since $F_\nu \propto \lambda^{-2}$ on the Rayleigh-Jeans tail, we can extrapolate Planet Nine's flux density to these longer wavelengths:
\begin{equation}
F_{\rm 150GHz} = 8~{\rm mJy} \left(\frac{T_{\rm eff}}{40~{\rm K}}\right) \left(\frac{R_p}{R_{\rm N}}\right)^2 \left(\frac{d}{700~{\rm AU}}\right) ^{-2}.
\end{equation}
 
\section{Proper Motion of Planet Nine}
Solar System objects move in the sky relative to the Earth due to both their own and Earth's orbital motion.

Neglecting Earth's orbital motion, the proper motion of a Solar System body on a circular orbit is $\mu_{\rm orb} = 2\pi/P$, where $P$ is the object's orbital period.  Using Kepler's Third Law, we get 
\begin{equation}
\mu_{\rm orb} = 0.3~{\rm arcsec/day}\left(\frac{d}{700~{\rm AU}} \right)^{-3/2}
\end{equation}
or, in the units of Kuiper Belt aficionados,
\begin{equation}
\mu_{\rm orb} = 10^{-2}~{\rm arcsec/hr}\left(\frac{d}{700~{\rm AU}} \right)^{-3/2}.
\end{equation}
Given the eccentricity range predicted by \cite{Batygin_2016}, the instantaneous orbital motion of planet nine could be a factor of 2--3 greater than this estimate.

Neglecting an object's orbital motion, its annual parallax as seen from Earth is $p = (d/{\rm pc})^{-1}$~arcseconds.  For Planet Nine, this gives a parallax of
\begin{equation}
p = 5~{\rm arcmin}  \left( \frac{d}{700~{\rm AU}}\right)^{-1}.
\end{equation}

The amplitude of the parallax motion seen from Earth is roughly 2$p$ per 6~months, or 
\begin{equation}
\mu_{\rm par} = 3.4~{\rm arcsec/day} \left( \frac{d}{700~{\rm AU}}\right)^{-1}
\end{equation}  
or
\begin{equation}
\mu_{\rm par} = 10^{-1}~{\rm arcsec/hr}\left(\frac{d}{700~{\rm AU}} \right)^{-1}.
\end{equation}
The instantaneous motion can be somewhat greater than this at opposition. Note, however, that mm telescopes exhibit no preference for opposition since they can operate during the day.  We therefore expect Planet Nine to move a few arcseconds per day, mostly due to parallax.   

\section{Detecting Planet Nine with CMB Experiments}
\begin{figure*}[htb]
\begin{center}
      \includegraphics[width=120mm]{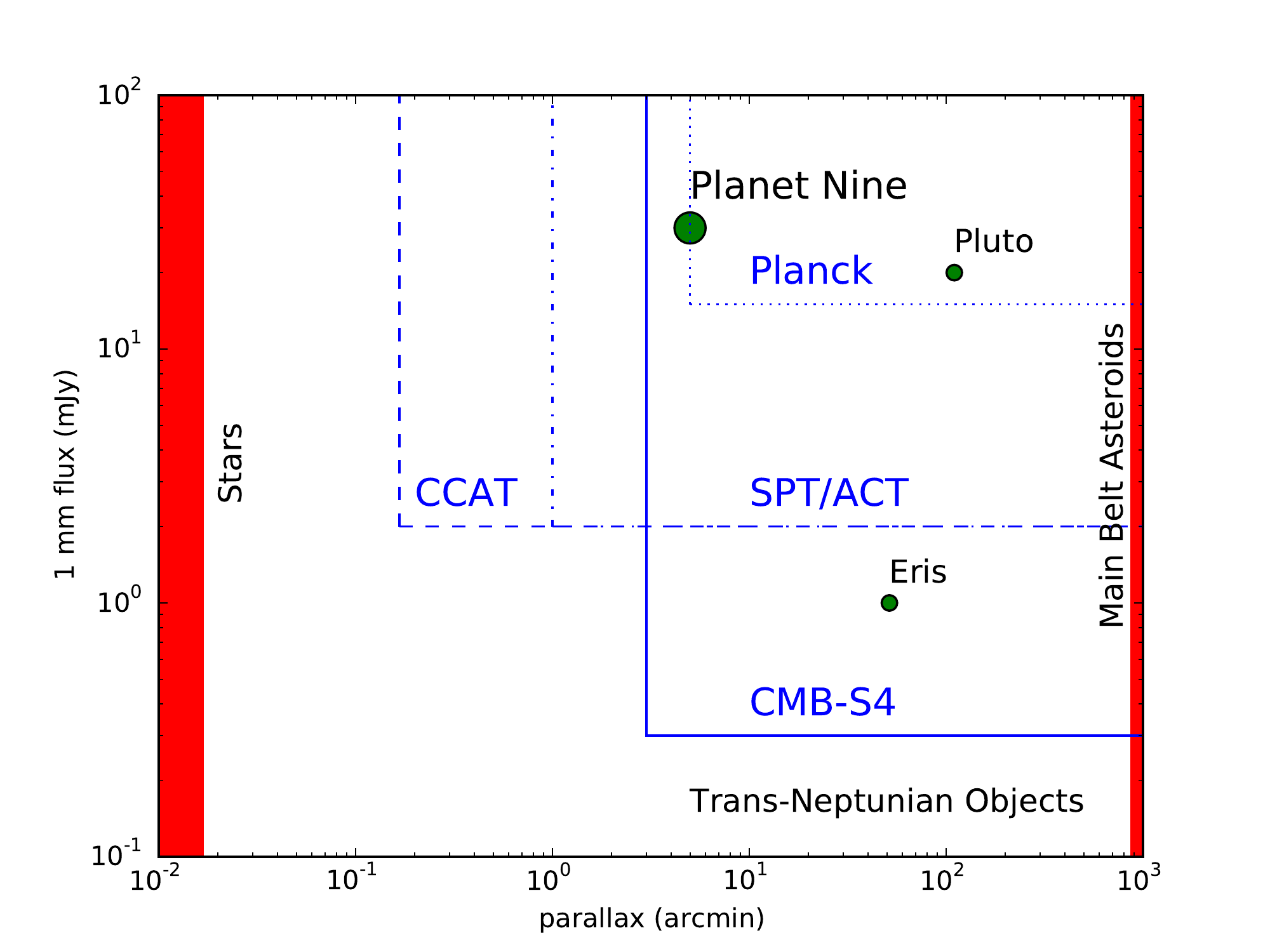}
    \caption{The search space for Planet Nine at mm wavelengths.  There is no other object with both parallax and mm flux within an order of magnitude of Planet Nine.  Stars are mostly fainter and all move much slower.  There are roughly 4000 asteroids at least as bright as Planet Nine, and many more that are fainter, but they all move orders of magnitude faster.  The largest Trans-Neptunian Objects, Pluto and Eris, are most similar to Planet Nine, but still move $>$10$\times$ faster.  The blue lines show the angular resolution and point-source sensitivity of current and planned CMB experiments.  The sensitivities are approximate because of the unknown scan strategy of future experiments, and because the reported numbers are for stationary sources: a moving source is less susceptible to confusion noise, but will require difference imaging.  Only experiments that can detect Planet Nine's mm flux and resolve its parallax motion can hope to discover it---this rules out Planck, unless Planet Nine is closer and brighter than we have assumed here.  Of the cosmology experiments that could identify the planet, those with greater sky coverage are better bets.  This tips the scales in favor of Advanced ACTPol or CMB-S4. \label{money_plot}}
    \end{center}
\end{figure*}

In order to find Planet Nine, a mm telescope needs sufficient point-source sensitivity to detect the object, and sufficient angular resolution to see its parallactic motion. The sensitivity of mm telescopes depends not only on their collecting area, but also the number of detectors. The angular resolution of mm telescopes, on the other hand, is simply related to their diameter, since they are nearly diffraction limited in most cases.  In other words, a small telescope with cutting-edge detectors may be sensitive to the mm flux of Planet Nine, but it is unlikely to resolve its sky motion. We now briefly discuss current and planned CMB experiments (also summarized in Figure~\ref{money_plot}).
  
Figure~4 of \citep{Fich_2014} suggests that the South Pole Telescope (SPT) and CCAT (with first generation instruments) have a 5$\sigma$ point-source detection limit of 1~mJy at 1~mm for a fiducial 1~hour scan of a 1 square degree field.  At that scan rate, a 1~year cosmological experiment could survey 365$\times$24 = 8760 square degrees, or 20\% of the sky.  This is large enough to put useful constraints on the location and luminosity of Planet Nine.  

Alternatively, the same instruments could survey 80\% of the sky in one year down to 2~mJy at 1~mm (in practice SPT can only access half the sky because it is located at the South Pole). The optimal strategy to search for Planet Nine would be to map the maximum area of the sky every few months in order to resolve its parallactic motion. For example, SPT could scan the entire Southern sky down to 4~mJy at 1~mm in under 3 months.      

If Planet Nine has a 1~mm flux of 30~mJy, then it should be robustly detected by SPT or CCAT within an hour using the fiducial scan strategy of 1 square degree per hour, or any of the other scan strategies described above. SCUBA-2 \citep{Thompson_2007} and JCMT \citep{Moore_2015}, however, are unlikely to be sensitive to Planet Nine unless it is significantly brighter than we predict.

SPT has an angular resolution of one arcminute \citep{Ruhl_2004}, so the parallax of Planet Nine is resolvable, and its motion is detectable with a few months baseline.  CCAT will have an angular resolution of ten arcseconds \citep{Fich_2014}, so the motion of Planet Nine would be easily detectable within a couple weeks. Planck has a resolution of 5~arcmin \citep{Cremonese_2002}, and can therefore just resolve the parallactic motion of Planet Nine.  The mission's point-source sensitivity, however, is insufficient to robustly detect Planet Nine unless it is hotter and/or closer than expected. 

Advanced ACTPol \citep{Henderson_2015} will observe 40\% of the sky and will have both the point source sensitivity and angular resolution to detect and identify Planet Nine.  Finally, CMB-S4 \citep{Abazajian_2013} will observe half of the sky with exquisite sensitivity, but it may only have an angular resolution of 3~arcmin, which would barely allow it to resolve Planet Nine's parallax. As a general rule, mm telescopes are nearly diffraction limited, so a 5~m telescope has $\lesssim$1~arcmin resolution at 1~mm; this is the smallest telescope that could convincingly resolve the parallax of Planet Nine if it is at 700~AU. The 12~arcmin beam of BICEP \citep{Ade_2015}, for example, is unable to resolve the parallax motion of Planet Nine.  

\section{False Positives}
Planet Nine will not be the only moving mm source. Here we consider false positives, some of which may be worthy targets in their own right.  Researchers have previously studied the use of CMB experiments to study Solar System objects, and the non-Gaussian noise that unresolved bodies contribute to mm maps \citep{Babich_2007, Babich_2009, Maris_2009}.
 
\subsection{Stars}
Even the stars with the greatest proper motions move orders of magnitude slower than Planet Nine: $\lesssim$0.03~arcsec/day. Stars are also radio quiet and hence much fainter than Planet Nine: there are only a handful of stars in the 10--100~mJy range at 1~mm \citep{Muller_2014}.

\subsection{Asteroids}
\cite{Redman_1995} reported the mm flux densities of asteroids at opposition. They find that the cumulative distribution of asteroids is roughly a power law down to 30~mJy, and they estimate that there are $4000$ asteroids at least this bright after accounting for incompleteness (a factor of 2).  By comparison with \cite{Terai_2013}, this roughly corresponds to all asteroids greater than 20~km in diameter. 

Most asteroids at any given moment are not at opposition. Since the nightsides of asteroids are cooler, they are less bright than the daysides. But mm-radiation originates from relatively deep in the asteroid, where the diurnal cycle is attenuated, so asteroids should have muted phase variations at these wavelengths. Asteroids at opposition are roughly a factor of two closer to Earth, however, and hence a factor of four brighter than asteroids that are 60$^\circ$ from the Sun.  Applying this factor to the steep power law in size and mm brightness, we expect 10$\times$ fewer asteroids at sky locations 60$^\circ$ from the Sun than at opposition. 

We treat the asteroid belt as being 10$^\circ$ wide and centered on the ecliptic \citep[a rough approximation of][]{Minton_2010}.  The asteroid belt therefore occupies approximately $360\times10= 3600$ square degrees on the sky.  At the SPT angular resolution of 1~arcmin, the asteroid belt therefore spans $60\times60\times3600 \approx 1.3\times10^7$ resolution elements.  We therefore expect an $F_\nu > 30$~mJy asteroid per every few thousand resolution elements, so we are far from the confusion-limit.    

For the most distant Main Belt asteroids \citep[4~AU;][]{Minton_2010}, we estimate proper motions of $\mu_{\rm orb} \approx 12$~arcmin/day and $\mu_{\rm par} \approx 10$~arcmin/day.  In other words, although there will be thousands of asteroids comparably bright to Planet Nine at mm wavelengths, they will move orders of magnitude faster, and their motion in the sky will be roughly equal parts proper motion and parallax.  

\subsection{Trans-Neptunian Objects}
Like Planet Nine, Trans-Neptunian Objects (TNOs) move slowly, mostly due to parallax.  Fortunately, only the very brightest of these objects would have comparable mm flux to Planet Nine: Pluto is 10--20~mJy \citep{Greaves_2015}, while Eris is about 1~mJy \citep{Bertoldi_2006}. But even here, TNOs betray themselves by their relatively large parallax.  The only way for an object to have Planet Nine's parallax is for it to be at Planet Nine's distance from the Sun.  And at those distances, only a bona fide planet could have a 1~mm flux density of 30~mJy.  In other words, the only false positive for Planet Nine is \emph{another} planet, which would be an equally exciting discovery.

\section{Discussion}
In order to detect Planet Nine, one needs to construct a mm map of the sky every few months and search for a $\sim$30~mJy source with annual parallax of $\sim$5~arcmin (Figure~\ref{money_plot}).  This is within reach of many current and planned cosmology experiments, provided they map large swaths of the sky and return to the same regions every few months.

This is a mode of operation that is easily accessible---but not guaranteed---for cosmology surveys. Some experiments will tile smaller fields and then move on, coming back once in a while or never.   Others survey large swaths over and over---this strategy is ideally suited for piggybacking Solar System science. Even in the worst case scenario of one and done, it takes a month or two to get to survey depth on a few hundred square degrees, during which the parallax motion of Planet Nine might betray its presence in a datacube.

Difference imaging at mm wavelengths will reveal stationary flickering AGN \citep[e.g.,][]{Hobbs_1977} and rapidly moving asteroids. Critically, asteroids move orders of magnitude faster on the sky than Planet Nine.  If Planet Nine is as radio bright as we predict, then there are sufficiently few foreground asteroids that they can be spatially resolved and tracked with hourly observations. Note that CMB experiments need not ``discover'' these asteroids: any object as mm-bright as 30~mJy is already in the Minor Planet Center (MPC) archive.\footnote{www.minorplanetcenter.net/iau/lists/ArchiveStatistics.html}

Current and planned cosmology experiments could observe thousands to millions of asteroids at mm wavelengths, constraining their vertical and horizontal temperature gradients, shapes, and albedo markings.  These experiments could also discover Planet Nine and measure its effective temperature.  Using Uranus and Neptune to bracket plausible internal heat fluxes, we expect Planet Nine to have an effective temperature of 30--50~K. Although we have adopted 40~K as our fiducial value throughout this Letter, our conclusions are essentially unchanged for any temperature in that range.  More importantly, estimating the actual effective temperature of Planet Nine by measuring its mm flux would help inform interior models of Uranus, Neptune, and possibly sub-Neptune exoplanets.

\subsection{What if Planet Nine is Fainter?}
If our assumptions about Planet Nine are wildly off then it could be much fainter. Lack of any internal heat source would lead to an effective temperature of $\sim$10~K if it has a similar Bond albedo to Uranus and Neptune, while a rocky planet of the same mass would be smaller.  Lastly, if Planet Nine lies at 1000 instead of 700~AU, then its flux would be halved.  If we conservatively adopt all of these scenarios, then the 1~mm flux density of Planet Nine could be as low as 1~mJy. This pessimistic scenario does not change our basic conclusion: Planet Nine---and its parallax motion---would still be detectable with many mm telescopes. Nonetheless, the search would be more challenging. 

If Planet Nine is only a few mJy then there are many more false positives.  The brightness distribution of asteroids is steep: extrapolating the power law of \cite{Redman_1995} down to flux densities of 1~mJy yields $10^6$ asteroids.   About $7\times10^{5}$ of these asteroids are already in the MPC archives, of which roughly $5\times10^5$ have sufficiently precise ephemerides to predict in which arcmin$\times$arcmin resolution element they lie \citep{Tedesco_2002, Bottke_2005}. The Large Synoptic Sky Survey (LSST) will find and track all of these---and many more---within half a decade of operations \citep{Jones_2016}. Nonetheless, near-term CMB experiments with sensitivities of 1~mJy have the potential to significantly increase the number of known asteroids. 

If we adopt the same 10 degree wide asteroid belt as above, we find that next-generation mm telescopes with 10~arcsec resolution will not be confusion limited, even for a fainter Planet Nine. Current telescopes such as SPT, however, are at the confusion limit near the ecliptic plane: 1 asteroid for every 10 resolution elements \citep{Franceschini_1982}.    In this more daunting regime, there are two approaches for dealing with foreground asteroids: 1) treat them as noise that can be averaged over, or 2) identify and remove asteroids from the map.  We consider each strategy in turn. 

If $\sim$20~days of mm observations are combined into a single mm-map, the asteroids will have moved a few degrees over the course of the observations.  They will therefore appear as streaks in the map.  Subtracting two such maps with millions of streaks each will therefore produce a proverbial haystack of streaks (half negative and half positive) with millions of apparent point sources at the intersections of streaks (since no two asteroids have exactly the same mm flux).  This haystack is further complicated by the fact that the flux from individual asteroids varies in time due to rotational effects (non-spherical shape and non-uniform albedo).  It is not obvious how to identify a faint slowly-moving point-source in this mess. 

We instead advocate to search for Planet Nine in the RA-DEC-time data cube. Asteroids will move from one resolution element to the next in approximately 2~hours.  By stacking images taken within an hour of each other, and comparing this hourly average to the following hourly average, asteroids can be identified.  Planet Nine will not move appreciably on this timescale, but will show up when comparing images taken weeks--months apart.  This seems like a surmountable problem of matched filtering that should be studied in more detail.  

If Planet Nine is smaller and/or more distant than we have assumed, then it will be harder to detect with mm telescopes, but doubly so at optical and near-IR wavelengths, given the $1/d^4$ dependence of reflected sunlight.  A search at mm wavelengths is only hopeless if Planet Nine has no internal heat \emph{and} has an albedo much greater than 50\%; such an ice-ball would be easier to discover via an optical search. Fortunately, cosmologists have designed optical all-sky surveys that may be up to the task \citep{LSST_2009}.  

\acknowledgements
We thank Jonathan Fortney, \'Etienne Artigau, Tracy Webb, Matt Dobbs, and Alan Fitzsimmons for timely and constructive feedback.


\begin{thebibliography}{}
\bibitem[{{Abazajian} {et~al.}(2013){Abazajian}, {Arnold}, {Austermann},
  {Benson}, {Bischoff}, {Bock}, {Bond}, {Borrill}, {Calabrese}, {Carlstrom},
  {Carvalho}, {Chang}, {Chiang}, {Church}, {Cooray}, {Crawford}, {Dawson},
  {Das}, {Devlin}, {Dobbs}, {Dodelson}, {Dore}, {Dunkley}, {Errard}, {Fraisse},
  {Gallicchio}, {Halverson}, {Hanany}, {Hildebrandt}, {Hincks}, {Hlozek},
  {Holder}, {Holzapfel}, {Honscheid}, {Hu}, {Hubmayr}, {Irwin}, {Jones},
  {Kamionkowski}, {Keating}, {Keisler}, {Knox}, {Komatsu}, {Kovac}, {Kuo},
  {Lawrence}, {Lee}, {Leitch}, {Linder}, {Lubin}, {McMahon}, {Miller},
  {Newburgh}, {Niemack}, {Nguyen}, {Nguyen}, {Page}, {Pryke}, {Reichardt},
  {Ruhl}, {Sehgal}, {Seljak}, {Sievers}, {Silverstein}, {Slosar}, {Smith},
  {Spergel}, {Staggs}, {Stark}, {Stompor}, {Vieregg}, {Wang}, {Watson},
  {Wollack}, {Wu}, {Yoon}, \& {Zahn}}]{Abazajian_2013}
{Abazajian}, K.~N., {Arnold}, K., {Austermann}, J., {et~al.} 2013, ArXiv
  e-prints, arXiv:1309.5383

\bibitem[Babich et al.(2007)]{Babich_2007} Babich, D., Blake, C.~H., \& Steinhardt, C.~L.\ 2007, \apj, 669, 1406

\bibitem[Babich \& Loeb(2009)]{Babich_2009} Babich, D., \& Loeb, A.\ 2009, New Astronomy, 14, 166

\bibitem[{{Batygin} \& {Brown}(2016)}]{Batygin_2016}
{Batygin}, K., \& {Brown}, M.~E. 2016, \aj, 151, 22

\bibitem[{{Bertoldi} {et~al.}(2006){Bertoldi}, {Altenhoff}, {Weiss}, {Menten},
  \& {Thum}}]{Bertoldi_2006}
{Bertoldi}, F., {Altenhoff}, W., {Weiss}, A., {Menten}, K.~M., \& {Thum}, C.
  2006, \nat, 439, 563

\bibitem[{{BICEP2 and Keck Array Collaborations} {et~al.}(2015){BICEP2 and Keck
  Array Collaborations}, {Ade}, {Aikin}, {Barkats}, {Benton}, {Bischoff},
  {Bock}, {Bradford}, {Brevik}, {Buder}, {Bullock}, {Dowell}, {Duband},
  {Filippini}, {Fliescher}, {Golwala}, {Halpern}, {Hasselfield}, {Hildebrandt},
  {Hilton}, {Hui}, {Irwin}, {Kang}, {Karkare}, {Kaufman}, {Keating}, {Kefeli},
  {Kernasovskiy}, {Kovac}, {Kuo}, {Leitch}, {Lueker}, {Megerian},
  {Netterfield}, {Nguyen}, {O'Brient}, {Ogburn}, {Orlando}, {Pryke}, {Richter},
  {Schwarz}, {Sheehy}, {Staniszewski}, {Sudiwala}, {Teply}, {Thompson},
  {Tolan}, {Turner}, {Vieregg}, {Weber}, {Wong}, \& {Yoon}}]{Ade_2015}
{BICEP2 and Keck Array Collaborations}, {Ade}, P.~A.~R., {Aikin}, R.~W.,
  {et~al.} 2015, \apj, 806, 206

\bibitem[Bottke et al.(2005)]{Bottke_2005} Bottke, W.~F., Durda, D.~D., Nesvorn{\'y}, D., et al.\ 2005, \icarus, 175, 111 

\bibitem[Brown \& Batygin(2016)]{Brown_2016} Brown, M.~E., \& Batygin, K.\ 2016, arXiv:1603.05712 

\bibitem[{{Cremonese} {et~al.}(2002){Cremonese}, {Marzari}, {Burigana}, \&
  {Maris}}]{Cremonese_2002}
{Cremonese}, G., {Marzari}, F., {Burigana}, C., \& {Maris}, M. 2002, New
  Astronomy, 7, 483

\bibitem[{{Fich}(2014)}]{Fich_2014}
{Fich}, M., and the CCAT Team, 2014, White Paper for Midterm Review, \verb+http://casca.ca/wp-content/uploads/2014/09/CCAT_WP_MTR.pdf+

\bibitem[Fienga et al.(2016)]{Fienga_2016} Fienga, A., Laskar, J., Manche, H., \& Gastineau, M.\ 2016, \aap, 587, L8 

\bibitem[Franceschini(1982)]{Franceschini_1982} Franceschini, A.\ 1982, \apss, 86, 3 

\bibitem[de la Fuente Marcos \& de la Fuente Marcos(2014)]{Fuente_Marcos_2014} de la Fuente Marcos, C., \& de la Fuente Marcos, R.\ 2014, \mnras, 443, L59 

\bibitem[de la Fuente Marcos et al.(2015)]{Fuente_Marcos_2015} de la Fuente Marcos, C., de la Fuente Marcos, R., \& Aarseth, S.~J.\ 2015, \mnras, 446, 1867 

\bibitem[de la Fuente Marcos \& de la Fuente Marcos(2016)]{Fuente_Marcos_2016} de la Fuente Marcos, C., \& de la Fuente Marcos, R.\ 2016, \mnras, accepted, arXiv:1603.06520 

\bibitem[Ginzburg et al.(2016)]{Ginzburg_2016} Ginzburg, S., Sari, R., \& Loeb, A.\ 2016, arXiv:1603.02876 

\bibitem[{{Greaves} {et~al.}(2015){Greaves}, {Whitelaw}, \&
  {Bendo}}]{Greaves_2015}
{Greaves}, J.~S., {Whitelaw}, A.~C.~M., \& {Bendo}, G.~J. 2015, \mnras, 449,
  L82

\bibitem[{{Henderson} {et~al.}(2015){Henderson}, {Allison}, {Austermann},
  {Baildon}, {Battaglia}, {Beall}, {Becker}, {De Bernardis}, {Bond},
  {Calabrese}, {Choi}, {Coughlin}, {Crowley}, {Datta}, {Devlin}, {Duff},
  {Dunner}, {Dunkley}, {van Engelen}, {Gallardo}, {Grace}, {Hasselfield},
  {Hills}, {Hilton}, {Hincks}, {Hlozek}, {Ho}, {Hubmayr}, {Huffenberger},
  {Hughes}, {Irwin}, {Koopman}, {Kosowsky}, {Li}, {McMahon}, {Munson}, {Nati},
  {Newburgh}, {Niemack}, {Niraula}, {Page}, {Pappas}, {Salatino}, {Schillaci},
  {Schmitt}, {Sehgal}, {Sherwin}, {Sievers}, {Simon}, {Spergel}, {Staggs},
  {Stevens}, {Thornton}, {Van Lanen}, {Vavagiakis}, {Ward}, \&
  {Wollack}}]{Henderson_2015}
{Henderson}, S.~W., {Allison}, R., {Austermann}, J., {et~al.} 2015, ArXiv
  e-prints, arXiv:1510.02809

\bibitem[{{Hildebrand} {et~al.}(1985){Hildebrand}, {Loewenstein}, {Harper},
  {Orton}, {Keene}, \& {Whitcomb}}]{Hildebrand_1985}
{Hildebrand}, R.~H., {Loewenstein}, R.~F., {Harper}, D.~A., {et~al.} 1985,
  Icarus, 64, 64

\bibitem[Hobbs \& Dent(1977)]{Hobbs_1977} Hobbs, R.~W., \& Dent, W.~A.\ 1977, \aj, 82, 257

\bibitem[Jones et al.(2016)]{Jones_2016} Jones, R.~L., Juri{\'c}, M., \& Ivezi{\'c}, {\v Z}.\ 2016, IAU Symposium, 318, 282 

\bibitem[Linder \& Mordasini(2016)]{Linder_2016} Linder, E.~F., \& Mordasini, C.\ 2016, arXiv:1602.07465 

\bibitem[LSST Science Collaboration et al.(2009)]{LSST_2009} LSST Science Collaboration, Abell, P.~A., Allison, J., et al.\ 2009, arXiv:0912.0201 

\bibitem[Maris \& Burigana(2009)]{Maris_2009} Maris, M., \& Burigana, C.\ 2009, Earth Moon and Planets, 105, 81 

\bibitem[{{Minton} \& {Malhotra}(2010)}]{Minton_2010}
{Minton}, D.~A., \& {Malhotra}, R. 2010, Icarus, 207, 744

\bibitem[Moore et al.(2015)]{Moore_2015} Moore, T.~J.~T., Plume, 
R., Thompson, M.~A., et al.\ 2015, \mnras, 453, 4264

\bibitem[{{M{\"u}ller} {et~al.}(2014){M{\"u}ller}, {Balog}, {Nielbock}, {Lim},
  {Teyssier}, {Olberg}, {Klaas}, {Linz}, {Altieri}, {Pearson}, {Bendo}, \&
  {Vilenius}}]{Muller_2014}
{M{\"u}ller}, T., {Balog}, Z., {Nielbock}, M., {et~al.} 2014, Experimental
  Astronomy, 37, 253

\bibitem[{{Redman} {et~al.}(1995){Redman}, {Feldman}, {Pollanen}, {Balam}, \&
  {Tatum}}]{Redman_1995}
{Redman}, R.~O., {Feldman}, P.~A., {Pollanen}, M.~D., {Balam}, D.~D., \&
  {Tatum}, J.~B. 1995, \aj, 109, 2869

\bibitem[{{Ruhl} {et~al.}(2004){Ruhl}, {Ade}, {Carlstrom}, {Cho}, {Crawford},
  {Dobbs}, {Greer}, {Halverson}, {Holzapfel}, {Lanting}, {Lee}, {Leitch},
  {Leong}, {Lu}, {Lueker}, {Mehl}, {Meyer}, {Mohr}, {Padin}, {Plagge}, {Pryke},
  {Runyan}, {Schwan}, {Sharp}, {Spieler}, {Staniszewski}, \&
  {Stark}}]{Ruhl_2004}
{Ruhl}, J., {Ade}, P.~A.~R., {Carlstrom}, J.~E., {et~al.} 2004, in \procspie,
  Vol. 5498, Z-Spec: a broadband millimeter-wave grating spectrometer: design,
  construction, and first cryogenic measurements, ed. C.~M. {Bradford},
  P.~A.~R. {Ade}, J.~E. {Aguirre}, J.~J. {Bock}, M.~{Dragovan}, L.~{Duband},
  L.~{Earle}, J.~{Glenn}, H.~{Matsuhara}, B.~J. {Naylor}, H.~T. {Nguyen},
  M.~{Yun}, \& J.~{Zmuidzinas}, 11--29

\bibitem[Tedesco \& Desert(2002)]{Tedesco_2002} Tedesco, E.~F., \& Desert, F.-X.\ 2002, \aj, 123, 2070 

\bibitem[{{Terai} {et~al.}(2013){Terai}, {Takahashi}, \& {Itoh}}]{Terai_2013}
{Terai}, T., {Takahashi}, J., \& {Itoh}, Y. 2013, \aj, 146, 111

\bibitem[{{Taylor}(2010)}]{Taylor_2010}
{Taylor}, F. 2010, \emph{Planetary Atmospheres}, Oxford Press

\bibitem[Thompson et al.(2007)]{Thompson_2007} Thompson, M.~A., 
Serjeant, S., Jenness, T., et al.\ 2007, arXiv:0704.3202 

\bibitem[{{Trujillo} \& {Sheppard}(2014)}]{Trujillo_2014}
{Trujillo}, C.~A., \& {Sheppard}, S.~S. 2014, \nat, 507, 471

\end{thebibliography}
\end{document}